\documentclass[sigplan, nonacm]{acmart}

\usepackage{makecell}
\usepackage{tabularx}
\usepackage{subcaption}
\usepackage{siunitx}
\usepackage{cleveref}
\usepackage{hyperref}

\usepackage{listings}
\usepackage{xcolor}

\usepackage{placeins}

\lstdefinelanguage{yaml}{
  keywords={true,false,null},
  keywordstyle=\color{blue},
  comment=[l]{\#},
  commentstyle=\color{gray}\ttfamily,
  stringstyle=\color{orange},
  morestring=[b]',
  morestring=[b]"
}

\begin{document}

\title[GPU Memory and Utilization Estimation for Training-Aware Resource Management]{GPU Memory and Utilization Estimation for Training-Aware Resource Management: Opportunities and Limitations}

\author{Ehsan Yousefzadeh-Asl-Miandoab}
\affiliation{
  \institution{IT University of Copenhagen}
  \country{Denmark}
}

\author{Reza Karimzadeh}
\affiliation{
  \institution{University of Copenhagen}
  \country{Denmark}
  }

\author{Danyal Yorulmaz}
\affiliation{
  \institution{IT University of Copenhagen}
  \country{Denmark}
  }

\author{Bulat Ibragimov}
\affiliation{%
  \institution{University of Copenhagen}
  \country{Denmark}
}

\author{Pınar Tözün}
\affiliation{
 \institution{IT University of Copenhagen}
 \country{Denmark}
}

\renewcommand{\shortauthors}{Yousefzadeh-Asl-Miandoab et al.}

\begin{abstract}

Collocating deep learning training tasks improves GPU utilization but risks resource contention, severe slowdowns, and out-of-memory (OOM) failures. Accurate memory estimation is essential for robust collocation, and GPU utilization estimation — a key proxy for contention — enables interference-aware scheduling.

Existing GPU memory estimators span three paradigms -- analytical models, CPU-side libraries, and ML-based estimators -- each with distinct limitations: dependence on detailed model specifications, intrusive integration, poor generalization, and varying latency overhead. GPU heterogeneity further complicates estimation, as identical tasks can exhibit different memory footprints across hardware generations. GPU utilization remains comparatively understudied, further complicated by the non-additive utilization metrics and GPU heterogeneity.

We conduct a systematic analysis of representative memory estimators from each paradigm -- Horus~\cite{horus}, PyTorch FakeTensor~\cite{faketensor2025}, and our lightweight ML-based estimator -- evaluating accuracy, generalizability, and overhead. We construct a synthetic dataset spanning MLPs, CNNs, and Transformers with controlled architectural variations, and train MLP- and Transformer-based estimators for memory prediction, and experiment with utilization estimation. Our evaluation reveals key tradeoffs and validates estimators against real-world unseen models. Significant challenges remain: analytical models lack generalization and cannot easily be extended to new GPU architectures or accurately reflect memory optimization savings; CPU-side libraries impose intrusive integration overhead; and both analytical and ML-based estimators rely on model specifications or computation graphs, limiting generalization across diverse model architectures and GPU hardware variants. We release all datasets, tools, and artifacts to support further research.

\end{abstract}


\maketitle

\section{Introduction}
\label{sec:introduction}

Accurate GPU memory estimation is critical for robust resource management in deep learning workloads, enabling appropriate GPU selection and safe task collocation without exceeding memory limits \cite{robroek2023analysis}. Estimating related metrics such as GPU utilization, as a proxy for interference, and memory bandwidth can further optimize collocation decisions and reduce performance degradations because of resource contention. However, estimation introduces latency into the critical path of resource manager decision-making. To mitigate this, estimation must be fast enough to meet scheduling deadlines, offloaded to machine learning frameworks (e.g., PyTorch) during model development, or pre-computed ahead of time while tasks are still queued and before they reach the scheduler.

Deep learning training tasks are compute-intensive and demand substantial GPU memory based on model architecture, model size, input data size, and batch size, making efficient memory management essential. Maximizing GPU utilization is also necessary for energy efficiency \cite{AnalysisOfMultiTenantGPUClusters, gao2024empirical, MLaaS}. Modern frameworks such as PyTorch and TensorFlow~\cite{pytorch, tensorflow} employ optimization techniques like \textbf{activation reuse} and \textbf{dynamic allocation}, introducing unpredictability into memory usage patterns of training tasks. Consequently, simple estimation formulas based solely on numbers of parameters, gradients and activations often fail to produce accurate predictions. This complexity is further compounded by the fact that identical training tasks can exhibit markedly different memory footprints across different GPU hardware models \cite{9826092}, owing to differences in software stacks, and hardware architecture. Beyond memory, GPU utilization estimation faces distinct challenges: hardware heterogeneity causes the same task to exhibit different utilization across different GPU models, and utilization metrics are \textbf{non-additive}, complicating collocation decisions.

Existing GPU memory estimation approaches — spanning analytical methods~\cite{horus, gao2020estimating}, library-based CPU-side tools~\cite{faketensor2025, 10.1145/3721462.3770773}, and ML-based estimators~\cite{gao2023runtime} — suffer from over- or underestimation, dependence on detailed model specifications or computation graph, poor generalizability, or latency overhead on the critical scheduling/mapping path. GPU utilization estimation remains comparatively understudied, despite its use as a proxy for interference-aware collocation \cite{horus} — a role complicated by its non-additive nature across collocated workloads. Reactive and/or profiling-based collocation systems such as Gandiva~\cite{Gandiva}, Gavel~\cite{gavel}, and Orion~\cite{strati2024orion} each address interference management under different assumptions, yet suffer from practical limitations: Gandiva requires intrusive ML framework modifications, Gavel incurs profiling overhead from its throughput estimator across accelerator types and its fixed profiling window may be too short to capture the behavior of long-running jobs or unnecessarily costly for short ones, and Orion requires costly offline per-kernel profiling.


In this paper, we analyze opportunities and limitations of the existing GPU memory and utilization estimators.
%
Our contributions are as follows:
\begin{list}{\labelitemi}{\leftmargin=1.5em}

\item{We survey existing GPU memory estimation methods by categorizing them into three — analytical, library-based, and ML-based — and provide a comprehensive qualitative and quantitative comparison of representative techniques from each class: Horus~\cite{horus}, PyTorch FakeTensor~\cite{faketensor2025}, and our own \textbf{GPUMemNet}.}

\item{We introduce \textbf{GPUMemNet} and \textbf{GPUUtilNet}, deep learning-based estimators for GPU memory consumption and utilization metrics, respectively. GPU memory estimators achieve up to 97\% accuracy for MLPs and 82--87\% for CNN and Transformer models, with limited generalization to real-world architectures such as ResNet, EfficientNet, and BERT. While GPUUtilNet reveals the difficulty of learning the patterns in the model architecture-level data.

\item We release\footnote{All artifacts, datasets, and reproducibility scripts are publicly available at \url{https://github.com/itu-rad/GPUMemNet} and \url{https://github.com/itu-rad/GPUUtilNet}.} a fully reproducible estimation pipeline encompassing data generation scripts for MLP, CNN, and Transformer architectures, curated training datasets, and documented Jupyter notebooks, lowering the barrier for the community to extend and build upon our work.
}
\item{We provide a discussion of the key challenges, open limitations, and future directions in GPU memory and utilization estimation, with an emphasis on their implications for collocation and resource management in shared GPU clusters.}
\end{list}

The remainder of this paper is organized as follows: \Cref{sec:related_work} reviews literature on GPU resource estimation and interference mitigation. 
\Cref{sec:dataset} describes our ML-based GPU memory use and utilization estimators GPUMemNet and GPUUtilNet.
\Cref{sec:analysis} provides a performance analysis of a set of representative estimators. 
Finally, \Cref{sec:discussion_limitations} discusses limitations and future work, with concluding remarks in \Cref{sec:conclusion}.

\section{Related Work}
\label{sec:related_work}

GPU resource estimation and GPU resource management, specifically collocation management have received growing attention as shared GPU clusters suffer from underutilization \cite{AnalysisOfMultiTenantGPUClusters,gao2024empirical,MLaaS}. We organize the related literature along two axes: \textit{static estimation} approaches that predict memory and utilization prior to execution (\Cref{sec:related_work:static}), and \textit{online, reactive} approaches that manage interference dynamically during collocation (\Cref{sec:related_work:reactive}).

\subsection{GPU Memory and Utilization Estimation}
\label{sec:related_work:static}

GPU memory use estimation methods fall into three categories. 

\textbf{\textit{Analytical methods}} derive memory estimates from model structure using closed-form accounting. Horus~\cite{horus} and DNNMem~\cite{gao2020estimating} are representative of this class; both require detailed architectural specifications and exhibit limited adaptability to transformer-based models and multi-GPU configurations, with DNNMem further lacking a publicly available implementation. LLMem~\cite{kim2024llmem} extends the analytical paradigm to distributed LLM fine-tuning, achieving low error rates, but remains tightly coupled to transformer decoder architectures, precluding generalization to other model families. A common limitation across analytical methods is their heavy reliance on detailed model specifications or computation graphs, which must either be explicitly provided by the user or obtained through intrusive coupling with the ML framework, and their limited extensibility — adapting to new configurations, such as multi-GPU settings, often requires revisiting and redesigning the underlying formulations rather than straightforward extension.

\textbf{CPU-side Library-based tools} perform symbolic or formula-driven estimation without  requiring GPU execution. xMem \cite{10.1145/3721462.3770773} utilizes the CPU to execute the initial iterations of a target task for a brief period to gather the necessary data for its estimation process. During this CPU execution, it collects detailed execution traces, including memory events and operator calls, allowing it to estimate peak memory requirements without needing a GPU. However, xMem still requires partial CPU execution to collect memory traces, introducing overhead and raising representativeness concerns — in training workloads, peak memory may not manifest during the initial iterations due to optimizer state initialization, gradient accumulation, or warmup phases, potentially leading to underestimation.
PyTorch FakeTensor~\cite{faketensor2025} propagates tensor shapes symbolically to estimate memory, requiring either model source code or an object instantiated from a checkpoint, while DeepSpeed~\cite{deepspeed_memory_requirements} provides built-in estimation utilities scoped exclusively to ZeRO-stage configurations, requiring either a model object or explicit parameter counts; both fail to account for framework-level overhead such as caching allocator reservations and dynamic memory allocation patterns that only emerge during actual execution.

\textbf{\textit{Machine Learning (ML)-based}} approaches attempt to overcome the rigidity of analytical methods; DNNPerf~\cite{gao2023runtime} employs graph neural networks over the model's computation graph, offering fine-grained operator-level accuracy, but at the cost of complex framework integration, inference latency that scales with graph complexity, and generalization challenges to unseen operator types or graph structures. More broadly, ML-based approaches are hindered by the difficulty of building sufficiently representative and diverse training datasets, making training, evaluation, and adaptation to new model families or hardware configurations a significant overhead.

Despite spanning three distinct paradigms, all methods introduce estimation overhead of varying magnitude, which can burden the critical path of resource manager decision-making.

In contrast, GPU utilization estimation remains comparatively understudied; Horus~\cite{horus} is among the few works to address it, yet its reliance on a small, curated model dataset limits its applicability to the diverse workloads encountered in production clusters.

\subsection{Collocation and Interference Management}
\label{sec:related_work:reactive}

Online, reactive approaches have been proposed to manage interference that arises when multiple training jobs share a GPU. Gandiva~\cite{Gandiva} supports collocation through time-slicing and introspective profiling of mini-batch progress rates, reverting packing decisions when significant slowdown is detected; however, this requires tight co-design between the ML framework and the resource manager, introducing intrusiveness to framework and resource management abstraction layers. Gavel~\cite{gavel} adopts a round-based scheduling mechanism with a fixed round length to make heterogeneity-aware allocation decisions, but a static round granularity is inherently ill-suited for workloads with widely varying iteration times — too coarse for short jobs and too fine for long ones. Orion~\cite{strati2024orion} achieves fine-grained, kernel-level collocation on a single GPU by classifying kernels as compute- or memory-bound through extensive, and costly, offline profiling and intercepting kernel launches via an additional layer in the software stack.
A common thread across these systems is that they treat interference reactively, after collocation has already been established, rather than proactively avoiding it through a priori estimation. 


\section{GPUMemNet \& GPUUtilNet}
\label{sec:dataset}

To complement the estimation methods based on analytical formulas and libraries
and strengthen the ML-based methods (described in \Cref{sec:related_work:static}),
we first present a methodology to create a curated dataset and use it for ML-based estimators.
This is necessary before we quantitatively evaluate the representative estimation methods from each category in \Cref{sec:analysis}.
\Cref{subsec:scripts} first outlines the considerations and steps for constructing a valid dataset, then \Cref{subsec:classification_vs_regression} justifies the formulation of GPU memory and utilization estimation as a classification task. 
Finally, \Cref{sebsec:selected_features} describes the dataset and selected features for the MLP- and Transformer-based estimators.

\subsection{Steps for Data Curation}
\label{subsec:scripts}

\subsubsection{Considerations for Data Collection}
\label{subsubsec:dataset_consideration}

To create a dataset with representative model features, we advocate for the following guidelines for data collection.

\textbf{Architecture over model types.} 
To ensure long-term relevance, we focus on architecture families (e.g., CNNs, Transformers) rather than specific models, which quickly become outdated.
This means that the data collection scripts must create synthetic versions of the architecture families.

\textbf{Representative and realistic designs.} 
Synthetic models are generated with feasible architectures (e.g., avoiding excessively deep models) and realistic components such as dropout and batch normalization.

\textbf{Diverse and balanced coverage.} 
We uniformly sample across input features and include varied topologies (e.g., pyramid, hourglass) to promote generalization.

\textbf{Input/output variation.} 
Models include varying input and output dimensions to assess their impact on GPU memory usage.

\subsubsection{Scripts to Collect Data}
\label{subsubsec:dataset_scripts}

To systematically collect hardware utilization metrics while running the synthetic models created based on the considerations outlined above, we develop a suite of scripts tailored to three different neural network architecture families: \textbf{MLPs}, \textbf{CNNs}, and \textbf{Transformers}. For each architecture, two main scripts are implemented:

%

\textbf{Configurable Model Script} defines a configurable neural network, supporting a flexible range of hyperparameters and architectural topologies (e.g., pyramid, hourglass, etc.). It ensures that only valid model configurations---feasible in terms of depth, width, and architecture---are generated. This validation guarantees the practicality of the network in real-world settings.

\textbf{Launcher Script} orchestrates data collection by uniformly sampling hyperparameters from predefined ranges. It launches a fixed number of configurations per run, with each configuration representing a unique data point. The parameters can be tweaked within the corresponding yaml file. \Cref{lst:mlp-yaml-config} shows the yaml file for MLP data collection. After configuring the parameters, running the script with \texttt{python launcher\_mlp.py}, the data collection starts.


% ...

\begin{figure}[t]
\centering
\begin{minipage}{0.85\linewidth}
\begin{lstlisting}[language=yaml, basicstyle=\ttfamily, 
    numbers=left, backgroundcolor=\color{gray!5}, 
    frame=lines]
base_data_dir: "mlp_dataset"

# num of data points
num_random_configs: 3000 
input_size:
  min: 4
  max: 4096
output_size:
  min_ratio: 0.25
batch_size:
  min: 1
  max: 256
...
\end{lstlisting}
\end{minipage}
\caption{YAML configuration for MLP generation}
\label{lst:mlp-yaml-config}
\end{figure}





Each generated model is trained for a fixed, configurable duration of one minute, during which GPU and CPU resources are monitored via \texttt{nvidia-smi}\cite{nvidia-smi}, \texttt{dcgmi}\cite{dcgmi}, and \texttt{top}~\cite{ubuntu_manuals_top}.
All raw monitoring logs and the model's architectural summary are saved in dedicated per-configuration directories.
Tables \ref{tab:mlp_parameters}-\ref{tab:target_extracted_parameters}
(all in Appendix) describe the curated dataset and its key parameters across model families.

\subsubsection{Data Cleaning and Feature Extraction}

For each model architecture type, a specialized \textbf{data cleaner script} parses the corresponding logs and summary files. These scripts extract:
\begin{itemize}
    \item Model-specific features (e.g., number of different layers, total parameters, depth-width profiles)
    \item Runtime GPU metrics (e.g., peak memory usage, utilization)
    \item CPU metrics (e.g., CPU\%, MEM\%)
\end{itemize}

The cleaned datasets are saved in CSV format for downstream processing. While the primary metrics in this work are GPU memory usage and utilization (SMACT, SMOCC, DRAMA) \cite{yousefzadeh2023profiling}, the dataset includes broader set of features to support future research directions, such as energy consumption or latency prediction.

%
%
%
%
%

\subsection{Classification vs. Regression}
\label{subsec:classification_vs_regression}

\begin{figure}[t]
    \centering
    \includegraphics[width=\linewidth, trim={0 0 0 0.1cm},clip]{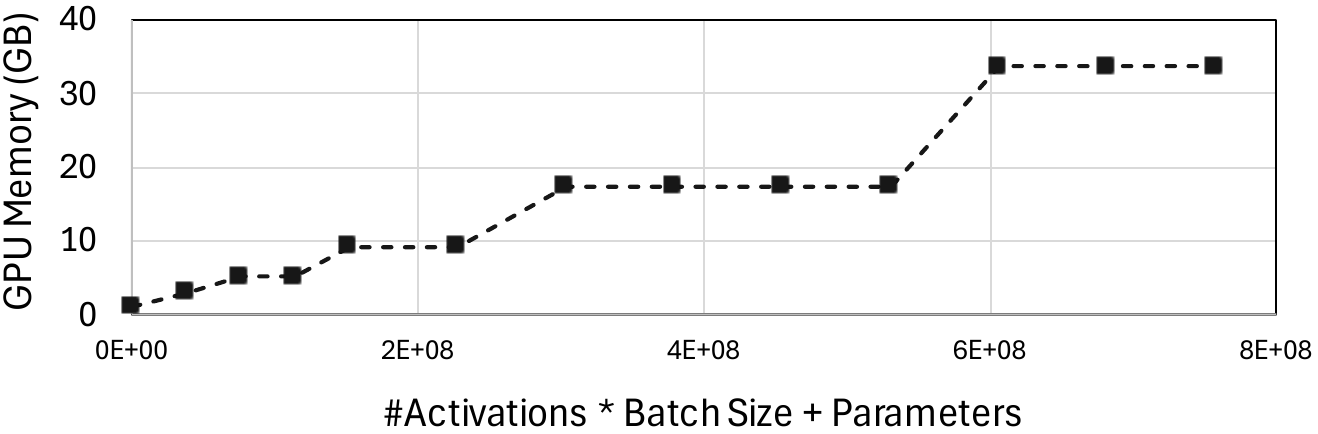}
    \caption{Staircase growth pattern for memory usage, MLPs on ImageNet \cite{ILSVRC15} and with batch\_size=32.}
    \label{fig:staricase_growth}
\end{figure}

As shown in \Cref{fig:staricase_growth},
GPU memory usage exhibits a staircase-like growth pattern.
This makes approaching the GPU memory use estimation as a regression challenging due to sharp discontinuities, flat plateaus with weak gradients, and spiking errors at step boundaries. These issues lead to instability and overfitting in regression models.

\begin{figure*}[t]
    \centering
    \begin{subfigure}[b]{0.32\textwidth}
        \centering
        \includegraphics[width=\linewidth, trim={0 0 0 0}, clip]{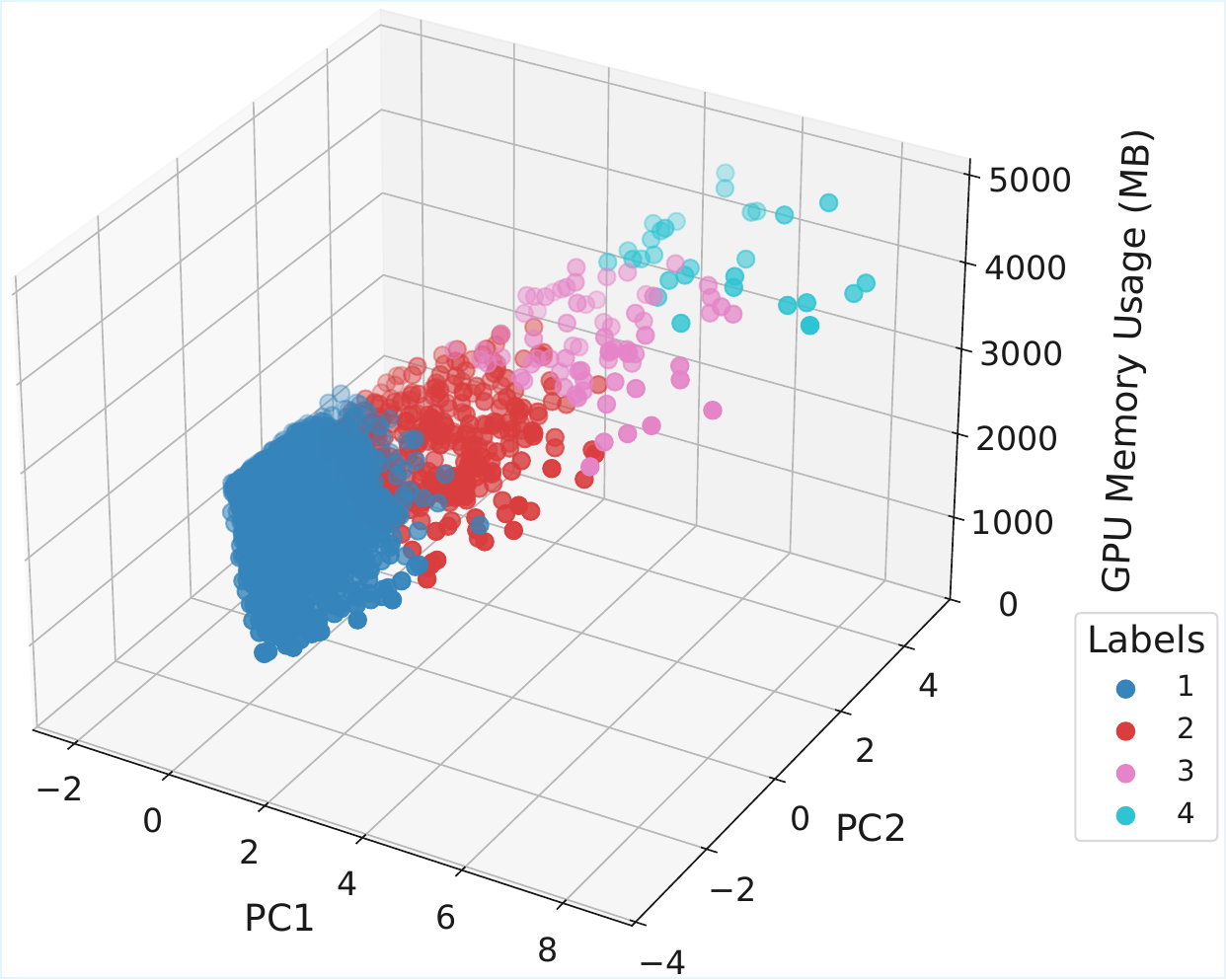}
        \caption{MLP}
        \label{fig:MLP_pca3d}
    \end{subfigure}
    \hfill
    \begin{subfigure}[b]{0.32\textwidth}
        \centering
        \includegraphics[width=\linewidth, trim={0 0 0 0}, clip]{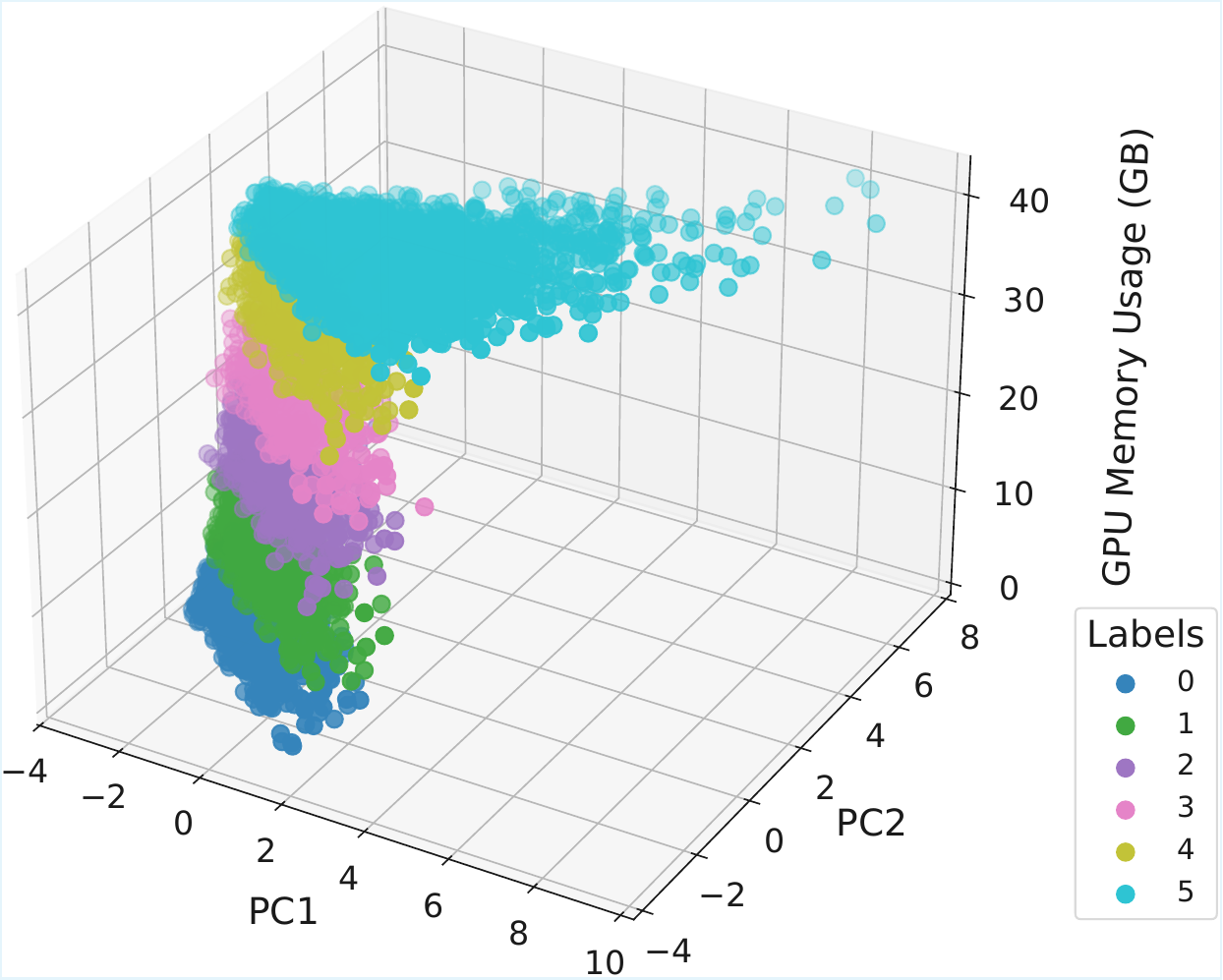}
        \caption{CNN}
        \label{fig:CNN_pca3d}
    \end{subfigure}
    \hfill
    \begin{subfigure}[b]{0.32\textwidth}
        \centering
        \includegraphics[width=\linewidth, trim={0 0 0 0}, clip]{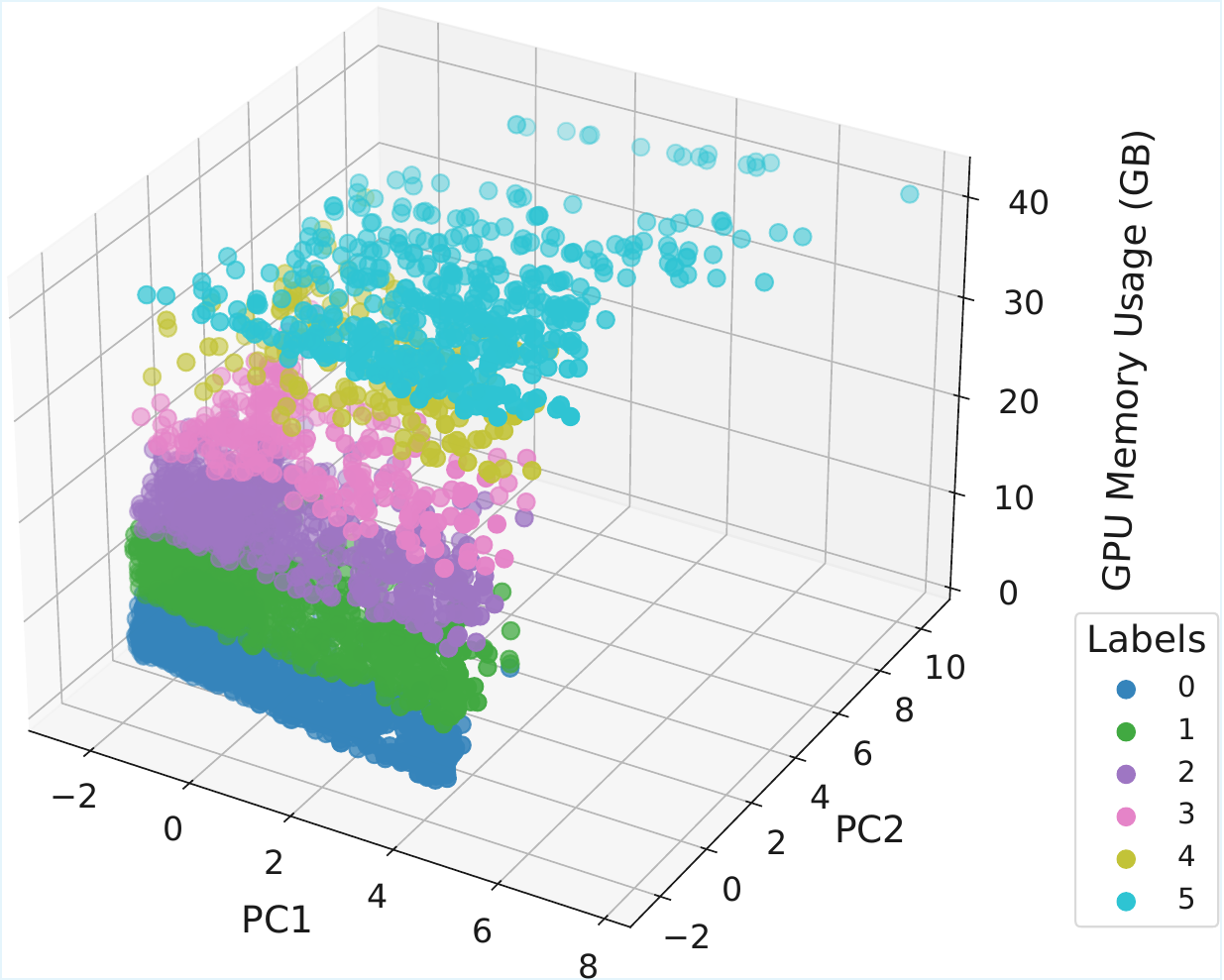}
        \caption{Transformer}
        \label{fig:TF_pca3d}
    \end{subfigure}

    \caption{Principal Component Analysis (PCA) of the dataset across different neural network architectures. The figure shows how discretizing the continuous GPU memory usage facilitates formulating the problem as a classification task.}
    \label{fig:dataset_PCA}
\end{figure*}

Therefore, we formulate the estimation task as classification by discretizing memory usage into fixed-size bins (e.g., 0--1 GB, 1--2 GB). Since CNNs and Transformers exhibit significantly higher memory demands than MLPs, larger bin sizes are used for these architectures to better capture their distribution. 

For estimation of the GPU utilization metrics, class imbalance and limited coverage of the metric range pose additional challenges, requiring bin adjustments. PCA visualization in  
\Cref{fig:dataset_PCA_util} (Appendix) further reveals that utilization metrics lack clear class separation. This stands in contrast to GPU memory, where more distinct class boundaries are observed in \Cref{fig:dataset_PCA}—specifically for MLPs—whereas the borders for CNNs and Transformers remain ambiguous, causing the model to become confused when learning these patterns.

\subsection{GPU Memory Need Estimation}
\label{sebsec:selected_features}



\begin{figure}[H]
    \centering
    \begin{subfigure}[b]{0.42\textwidth}
        \centering
    \includegraphics[width=0.7\linewidth, trim={0 0 0 0.1cm},clip]
    {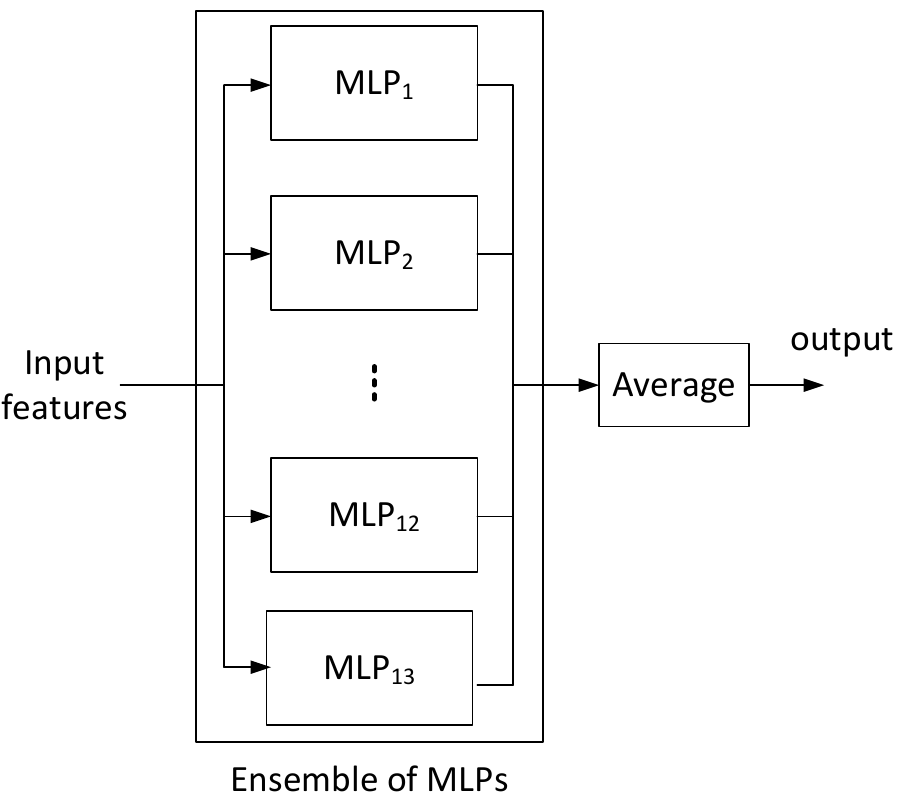}
        \caption{MLP Ensemble}
        \label{fig:mlp}
    \end{subfigure}
    \hfill
    \begin{subfigure}[b]{0.48\textwidth}
        \centering
        \includegraphics[width=0.7\linewidth, trim={0 0 0 0.1cm}, clip]{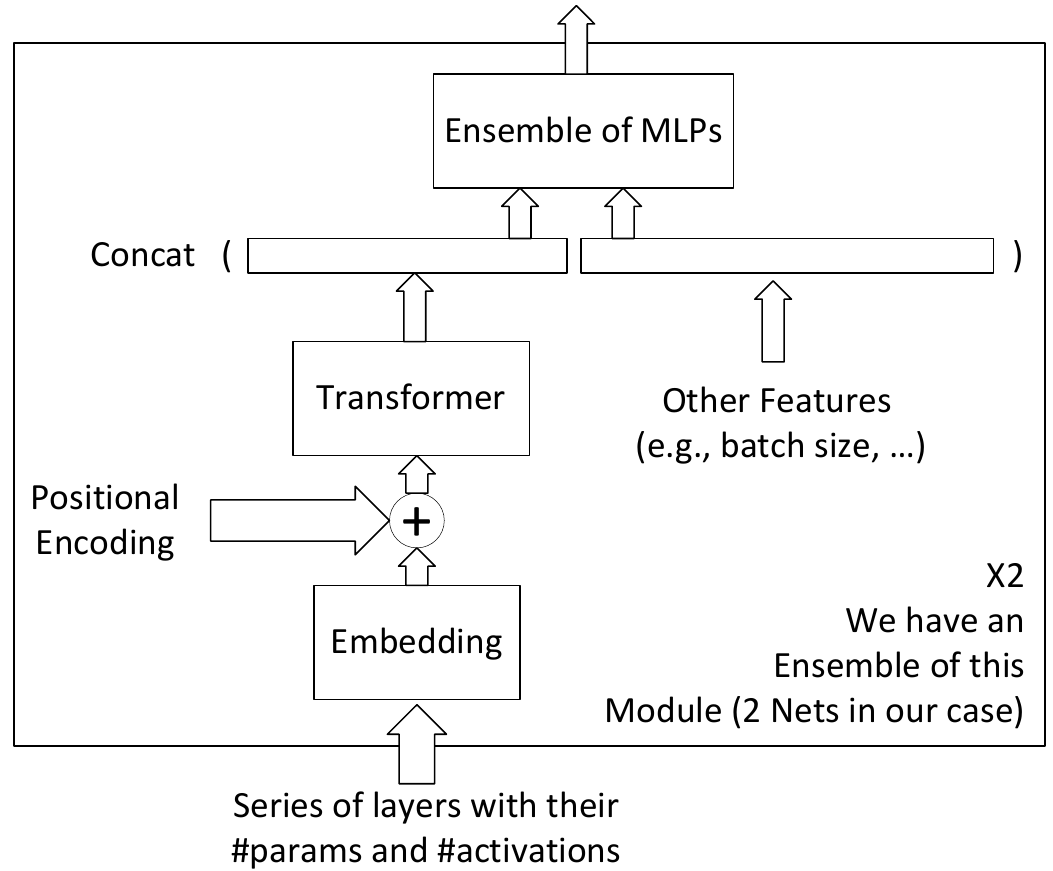}
        \caption{Transformer Ensemble.}
        \label{fig:transformer}
    \end{subfigure}
    \caption{GPUMemNet Estimator Model Architectures.}
    \label{fig:est_architecture}
\end{figure}

We implement and evaluate \textbf{ensemble-based estimators} for GPU memory usage prediction. In particular, we study two estimator families in GPUMemNet: an ensemble of MLPs (\Cref{fig:mlp}) and an ensemble of Transformer classifiers (\Cref{fig:transformer}). In both cases, the ensemble aggregates multiple base models with varied hyperparameters and structures to improve generalization, predictive accuracy, and training stability across different deep learning architecture families.

For the MLP-based estimator, each base model is a randomly structured feedforward network with 1 to 8 hidden layers and a progressively decreasing number of neurons per layer. The hidden dimension decays exponentially from a maximum of 8 to a minimum of 4. All hidden layers use ReLU activation and batch normalization. The final MLP ensemble prediction is obtained by averaging the outputs of multiple such base models.

For the Transformer-based estimator, we construct an ensemble of Transformer encoders with varying architectural configurations to better capture the sequential structure of neural network layers and their associated parameters and activations. Each Transformer classifier begins with an MLP-based input embedding layer, followed by positional encodings and a stack of Transformer encoder blocks. Each encoder configuration is defined by the embedding dimension $d$, the number of attention heads, the number of encoder layers, and the feedforward dimension. In our implementation, $d \in \{4,6\}$, the number of encoder layers ranges from 2 to 3, the number of attention heads is fixed to 1, and the feedforward dimension is fixed to 4. Dropout rates vary between 0.0 and 0.3. The encoded representation is then concatenated with structured auxiliary features and passed to a final MLP ensemble for classification.

All models are trained using cross-entropy loss and optimized with Adam. For evaluation, we use \textbf{3-fold stratified cross-validation} to ensure robust performance reporting. In each fold, 70\% of the split is used for training and 30\% for validation, while the test set is held out separately using an additional 30\% split. We evaluate both estimator families across different deep learning model architectures, including MLP- and Transformer-based workloads.

For all datasets (\Cref{subsubsec:dataset_scripts}), the MLP-based estimator uses: \textbf{batch size}, \textbf{number of parameters}, \textbf{number of activations}, \textbf{\#activations $\times$
batch size} and the
\textbf{number of architecture-specific layer types} (e.g., linear,
convolutional, batch normalization, and dropout layers). The MLP and CNN datasets
include \textbf{activation encoding (cos/sin)}, which is also added to the set of features for estimator training. 
The Transformer-based estimator extends all of the above with a \textbf{sequence of (layer type,
\#activations, \#parameters) tuples}. These features capture model scale and structure — the primary determinants of GPU memory requirements and utilization.

\Cref{tab:target_extracted_parameters} in Appendix presents the target variable ranges alongside total parameters and activations extracted from model summaries. GPU Memory estimator, \textbf{GPUMemNet}, test models are trained with cross-entropy loss (Adam optimizer) and evaluated via stratified 3-fold cross-validation (70\% train / 30\% validation per fold, with a held-out 30\% test set), reporting Accuracy and F1-score to measure overall correctness and class balance, respectively.

\subsection{GPU Utilization Estimator}
Given that GPU utilization metrics remain largely understudied compared to GPU memory estimation and the Horus dataset and models are currently unavailable, we train a deep learning-based estimator to predict three utilization metrics — SMACT, SMOCC, and DRAMA — using the dataset we curated for GPU memory estimation. These metrics, averaged over the execution lifetime of each training job, together capture how both the compute and memory units of the GPU are utilized, providing a more complete picture of resource usage beyond memory consumption alone. Following the same methodology applied to GPU memory estimation, we reuse the same input features and dataset, applying the same empirical evaluation approach to assess their predictive power.
\section{Quantitative Analysis of Estimators}
\label{sec:analysis}

This section quantitatively evaluates representative GPU memory and utilization estimators for training-aware resource management. We focus on three questions: how accurately these approaches estimate resource usage, how well they generalize across workloads and unseen models, and what practical overheads they impose. This analysis highlights the tradeoffs among analytical, CPU-side, and ML-based estimators.

\subsection{Experimental Environment}
\label{sec:analysis:env}

All data collection and experiments are conducted on an NVIDIA DGX Station A100 equipped with four NVIDIA A100 GPUs (40 GB HBM2 each), using PyTorch 2.4.1 and CUDA 12.2.

\subsection{Evaluated Estimators}
\label{sec:estimation:selected}

From the three classes of GPU memory need estimators surveyed in \Cref{sec:related_work:static}, we select one available method per paradigm for empirical evaluation, the Horus analytical formula~\cite{horus}, PyTorch FakeTensor~\cite{faketensor2025}, and our own lightweight ML-based estimator, \textbf{GPUMemNet}. 
The reasons for not including others are 
DNNMem \cite{gao2020estimating}, which is not publicly available, LLMem~\cite{kim2024llmem}, which is restricted to transformer decoder fine-tuning, and DeepSpeed's estimator~\cite{deepspeed_memory_requirements}, which is scoped exclusively to ZeRO-stage configurations and does not account for activation memory or framework overhead.
DNNPerf~\cite{gao2023runtime} relies on a GNN operating over the model's computation graph, which requires framework deserialization APIs to parse and reconstruct operator-level graph representations — a more complex and tightly coupled integration compared to a lightweight feature-based approach. Furthermore, its GNN-based inference latency scales with model and graph complexity, and constructing sufficiently representative training datasets for GNN-based estimators carries significant collection and curation overhead, motivating the design of our own lightweight estimator, \textbf{GPUMemNet}.

\subsection{Memory Estimators on Micro-benchmarks}
\label{sec:estimation:ubench}

\begin{figure}[t]
    \centering
    \includegraphics[width=\linewidth, trim={0 0 0 0.1cm},clip]{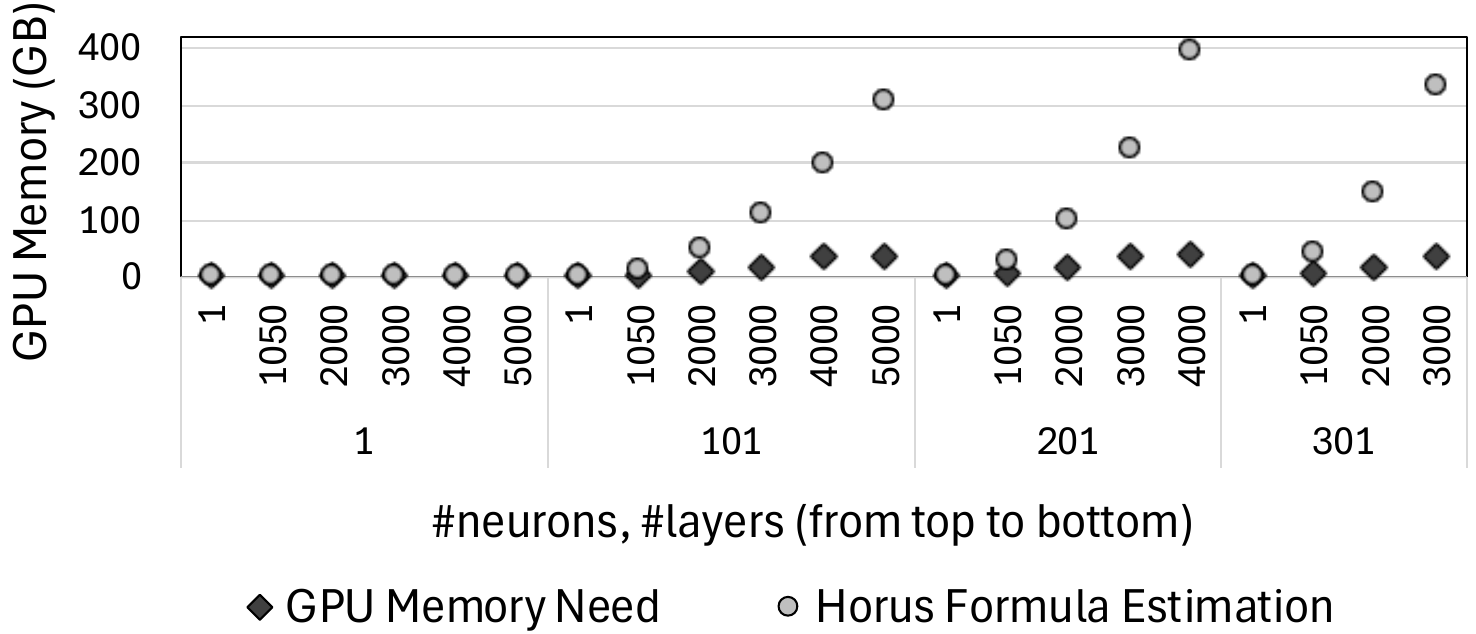}
    \caption{Actual GPU memory need vs Horus' estimations for MLP models with varying number of neurons and layers.}
    \label{fig:horus_misestimations}
\end{figure}

\textbf{Analytical.}
Horus~\cite{horus} estimates training memory via a lightweight analytical pass over model statistics, incurring negligible overhead. However, as \Cref{fig:horus_misestimations} demonstrates, it systematically overestimates memory consumption across MLP configurations due to its inability to account for framework-level optimizations such as activation reuse, dynamic memory management, and layer fusion. While this conservatism avoids OOM failures, it unnecessarily reduces collocation opportunities by over-reserving memory.

\textbf{CPU-side Libraries, tools.}
 PyTorch FakeTensor~\cite{faketensor2025} performs shape propagation — optionally symbolic — creating lightweight tensor surrogates that carry metadata (shape, dtype, device, and aliasing information) without allocating real memory, enabling analysis of model structure without modifying or running the training process itself. Evaluated across 2{,}030 model configurations from the \texttt{TIMM}~\cite{rw2019timm} library, FakeTensor consistently \emph{underestimates} peak memory usage (\Cref{tab:faketensor-acc-time}), as it does not account for the caching allocator's reserved memory, intermediate buffers, or optimizer states. A fixed safety margin (e.g., \textbf{+4,GB}) can partially compensate for this, providing a conservative bound suitable for collocation decisions. Its runtime overhead, while modest, is the most prominent among the three candidates and adds latency to the scheduling critical path. Additionally, FakeTensor fails on architectures that rely on unsupported operations, dynamic control flow, or CUDA-only kernels — including many Transformer variants, DLRM, and Mask R-CNN — limiting its practical coverage.


\begin{table}[t]
\caption{FakeTensor estimation accuracy and time over 2{,}030 training runs. Columns show percentiles; \% of runs with $>$8\,GB error is 0.49\% of the whole data.}
\centering
\begin{tabularx}{\columnwidth}{|l|c|c|c|c|c|>{\centering\arraybackslash}X|}
\hline
\textbf{Metric} & \textbf{P50} & \textbf{P80} & \textbf{P90} & \textbf{P95} & \textbf{P99} & \textbf{Max} \\
\hline \hline
\makecell[l]{Absolute\\error (GB)}
& 1.16 & 1.97 & 2.82 & 3.84 & 6.82 & $>$8\,GB \\
\hline
\makecell[l]{Estimation\\time (s)}
& 0.94 & 1.49 & 1.84 & 2.28 & 3.47 & 4.97 \\
\hline
\end{tabularx}
\label{tab:faketensor-acc-time}
\end{table}

\textbf{ML-based.}
GPUMemNet (\Cref{sec:dataset}) is trained on a curated synthetic dataset spanning MLP, CNN, and Transformer architecture families and formulates memory prediction as a classification task over discretized memory ranges. Unlike the analytical and library-based alternatives, it requires no model source code, no architectural specification beyond high-level configuration parameters, and introduces minimal inference latency.

\begin{figure*}[t]
    \centering  \includegraphics[width=\linewidth, trim={0 0 0 0.1cm},clip]{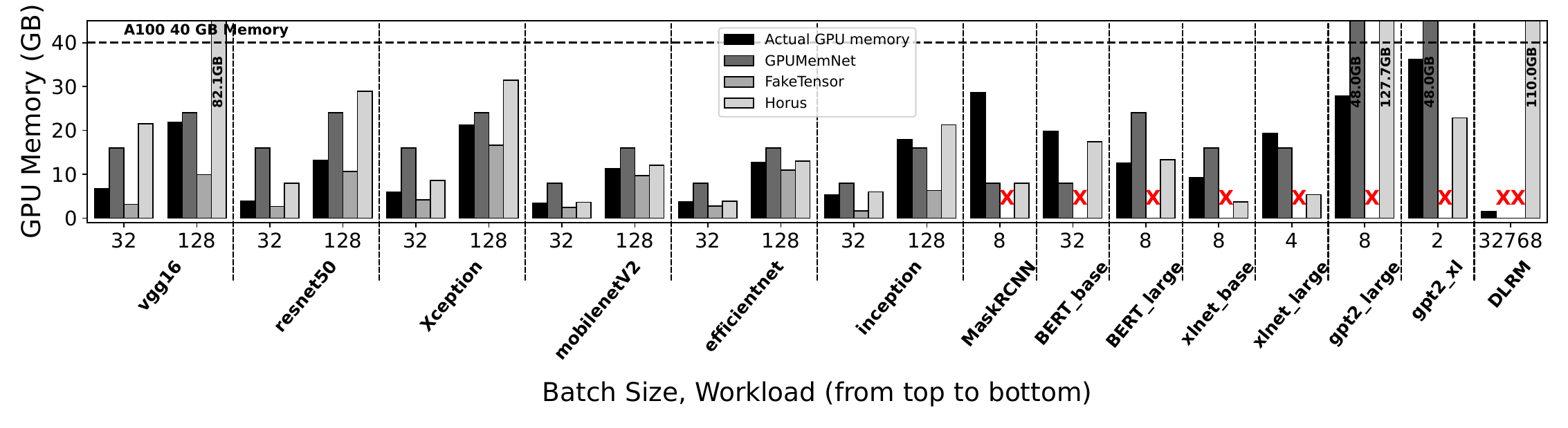}
    \caption{GPU memory estimation for real-world unseen CNN and Transformer models using Horus, FakeTensor, and GPUMemNet. FakeTensor fails at Transformer models and GPUMemNet cannot estimate for the unseen model, e.g., DLRM (denoted with \textcolor{red}{X}). GPUMemNet provides the closest estimations to actual GPU memory consumption and almost never underestimates.}
    \label{fig:GPUMemNet_eval1}
\end{figure*}

Results in \Cref{tbl:models_results} show that
GPUMemNet achieves high accuracy on MLPs (serving as a proof-of-concept) and strong performance on CNNs and Transformers (0.81–0.88 at 8\,GB memory range bins), supporting the classification formulation for practical memory estimation. Using an 8\,GB bin width preserves accuracy but can limit collocation opportunities, so we also experiment with smaller bins. For CNNs and Transformers, however, finer bins degrade performance --- reflecting complex, high-variance patterns that would require substantially \textbf{more data} to learn, incurring significant time and hardware costs (a core caveat of data-driven estimation).
A second challenge is drift: any change in frameworks or memory optimizations can invalidate labels, forcing full data recollection and reprocessing.
Third, generalization is brittle: ML estimators extrapolate poorly to unseen architectures or new layers. For example, probing a GPT-2–style model revealed 1D convolution layers—operators absent from GPUMemNet’s Transformer training set—explaining poor extrapolation. This underscores a practical need: keep the GPUMemNet datasets \emph{open} and continuously extended to cover new layers and variants, improving accuracy over time. Finally, the runtime cost of a GPUMemNet estimator model is low: at most \textbf{16\,ms} on an NVIDIA A100 (40\,GB) and \textbf{32\,ms} on an AMD EPYC CPU, measured over 100 runs per estimator.

\begin{table}[t]
\caption{Accuracy results for the GPU memory use estimations with MLP- and Transformer-based models.}
\centering
\begin{tabularx}{\columnwidth}{|l|X|c|c|c|}
\hline
\textbf{Dataset} & \textbf{Estimator} & \textbf{Bin} &
\textbf{Acc.} &
\textbf{F1-score} \\
\hline \hline
MLP 
    & \makecell[l]{MLP} 
    & 1GB & 0.95 & 0.93 \\ 
\cline{2-5}
    & \makecell[l]{MLP} 
    & 2GB & 0.97& 0.96 \\ 
\cline{2-5}
    & \makecell[l]{Transformer} 
    & 1GB & 0.97 & 0.96 \\ 
\cline{2-5} 
    & \makecell[l]{Transformer} 
    & 2GB & 0.98 & 0.97 \\ 
\hline

\makecell[l]{CNN}
    & \makecell[l]{MLP}
    & 8GB & 0.83 & 0.83 \\ 
\cline{2-5}
    & \makecell[l]{Transformer} 
    & 8GB & 0.81 & 0.81 \\ 
\hline

\makecell[l]{Transformer}
    & \makecell[l]{MLP} 
    & 8GB & 0.88 & 0.88 \\ 
\cline{2-5}
    & \makecell[l]{Transformer} 
    & 8GB & 0.86 & 0.86 \\ 
\hline

\end{tabularx}
\label{tbl:models_results}
\end{table}


\subsection{Memory Estimators on Real-World Models}
\label{sec:gpumemnet:results}

After the in-depth look at the individual strengths and weaknesses of the different GPU memory need estimation methods,
we compare all three estimators—the Horus formula \cite{horus}, FakeTensor \cite{faketensor2025}, and GPUMemNet—on diverse real-world models in \Cref{fig:GPUMemNet_eval1}.
We use GPUMemNet’s MLP-based estimators throughout, given their higher accuracy on CNNs and Transformers (\Cref{tbl:models_results}). The Horus formula can both under- or overestimate—underestimates risk OOM, while overestimates waste memory and reduce collocation. FakeTensor typically underestimates (risking OOM) and, for many Transformer models, produces \emph{no estimate} at all due to unsupported ops.
%
In contrast, GPUMemNet tends to \emph{overestimate} memory, reducing potential collocation gains. Its largest error appears on GPT-2—an out-of-distribution case for our training data—where unseen architectural elements drive the mismatch.
Furthermore, it cannot estimate for the DLRM as it has not been trained for it,
underlining the weakness of model-based estimation on unseen cases.

\subsection{Utilization Estimator}
\label{sec:gpumemnet:resultsutil}

\begin{figure}
    \centering
    \begin{subfigure}[b]{0.32\textwidth}
        \centering
        \includegraphics[width=\linewidth, trim={0 0 0 0}, clip]{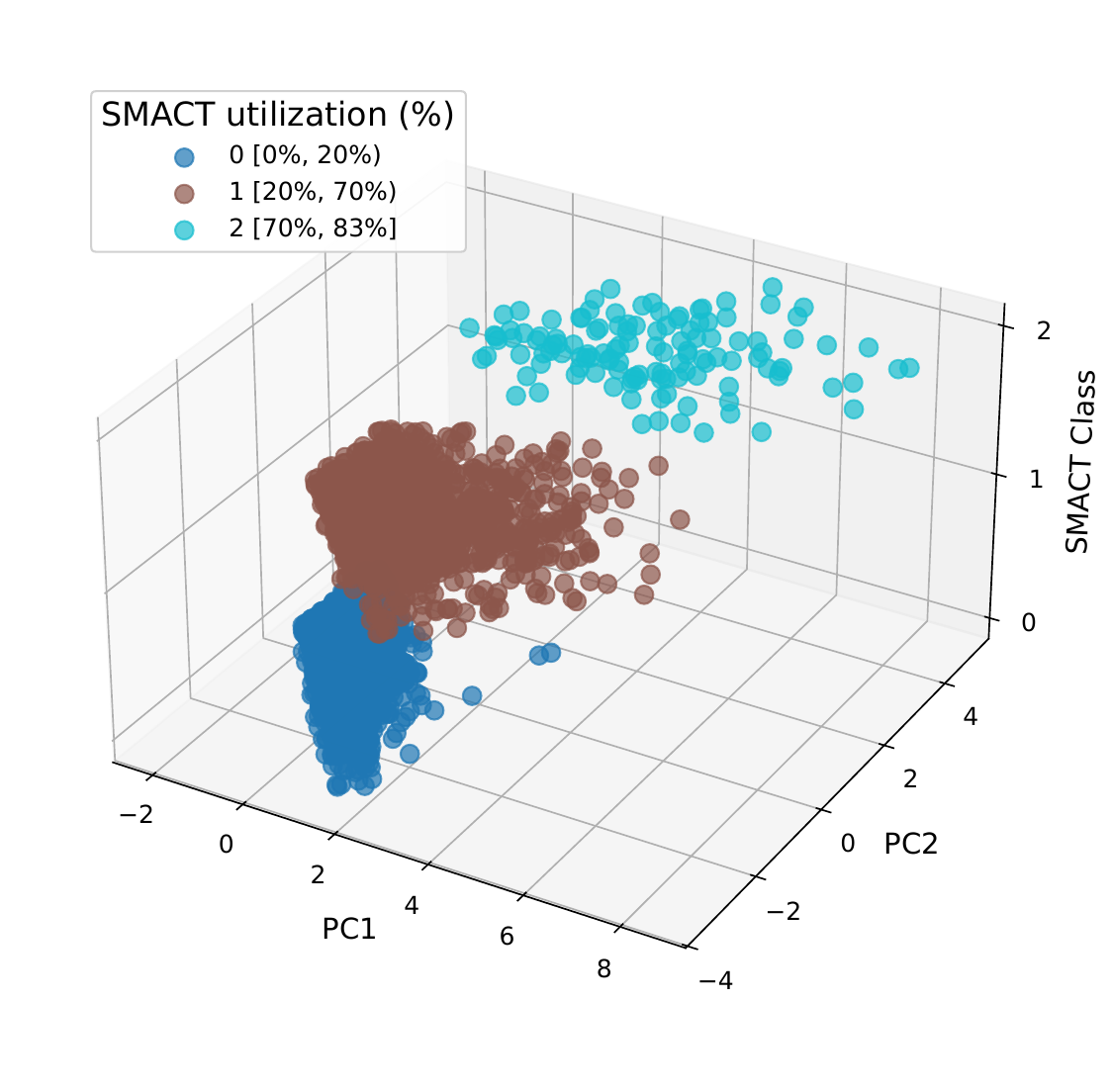}
        \caption{MLP SMACT}
        \label{fig:MLP_SMACT_pca3d}
    \end{subfigure}
    \begin{subfigure}[b]{0.32\textwidth}
        \centering
        \includegraphics[width=\linewidth, trim={0 0 0 0}, clip]{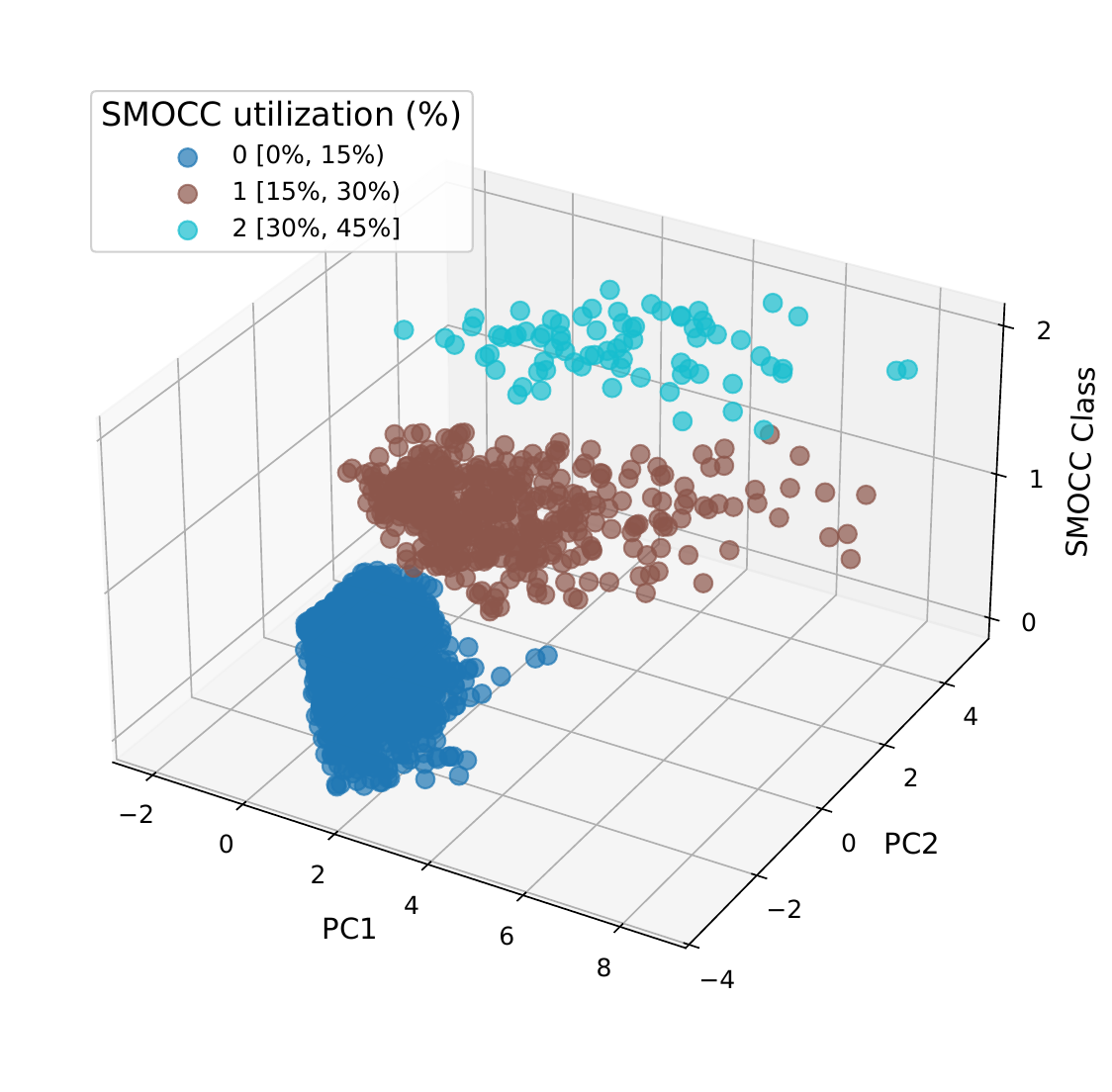}
        \caption{MLP SMOCC}
        \label{fig:MLP_SMOCC_pca3d}
    \end{subfigure}
    \begin{subfigure}[b]{0.32\textwidth}
        \centering
        \includegraphics[width=\linewidth, trim={0 0 0 0}, clip]{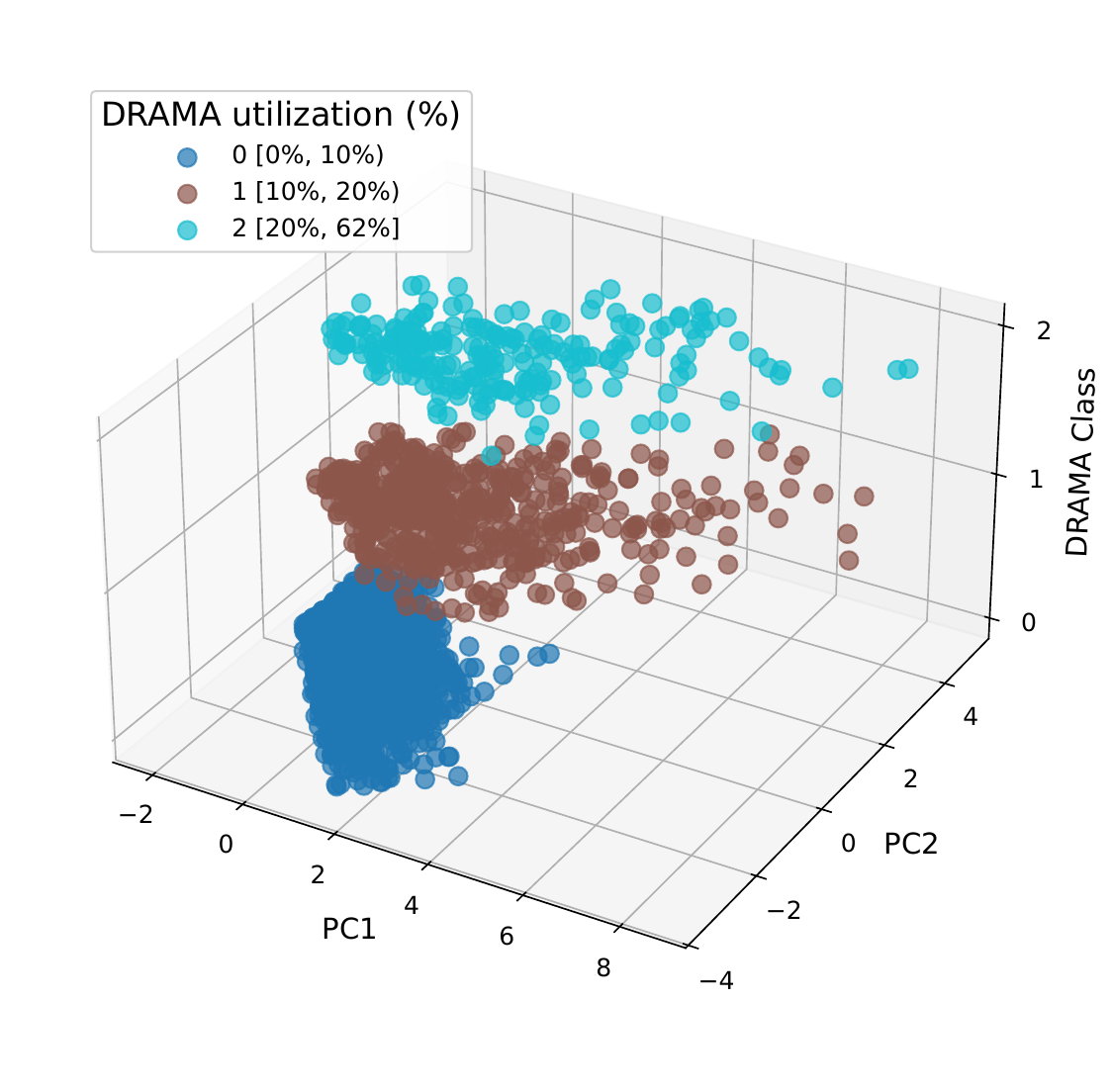}
        \caption{MLP DRAMA}
        \label{fig:MLP_DRAMA_pca3d}
    \end{subfigure}
    \begin{subfigure}[b]{0.32\textwidth}
        \centering
        \includegraphics[width=\linewidth, trim={0 0 0 0}, clip]{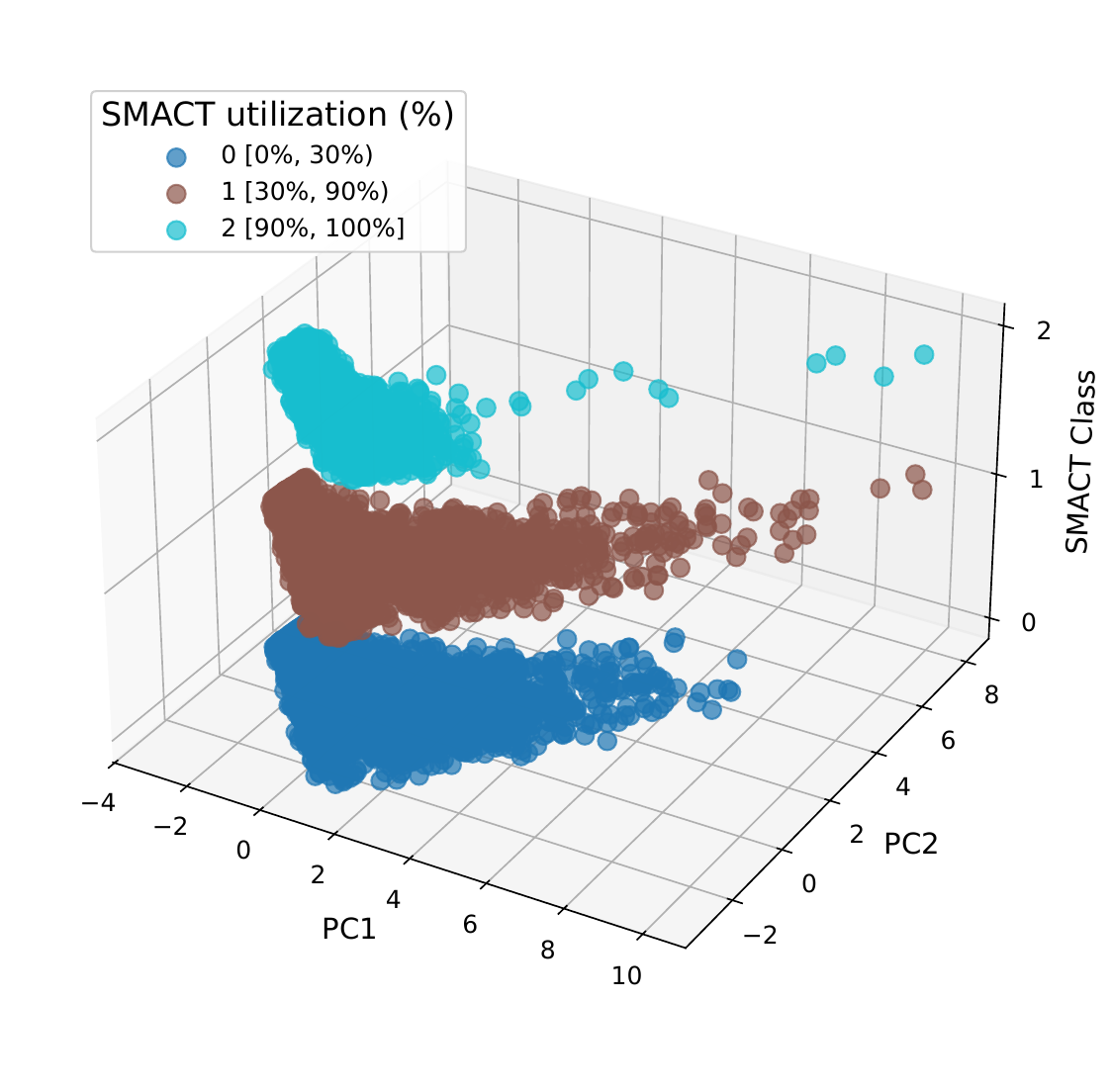}
        \caption{CNN SMACT}
        \label{fig:CNN_SMACT_pca3d}
    \end{subfigure}
    \begin{subfigure}[b]{0.32\textwidth}
        \centering
        \includegraphics[width=\linewidth, trim={0 0 0 0}, clip]{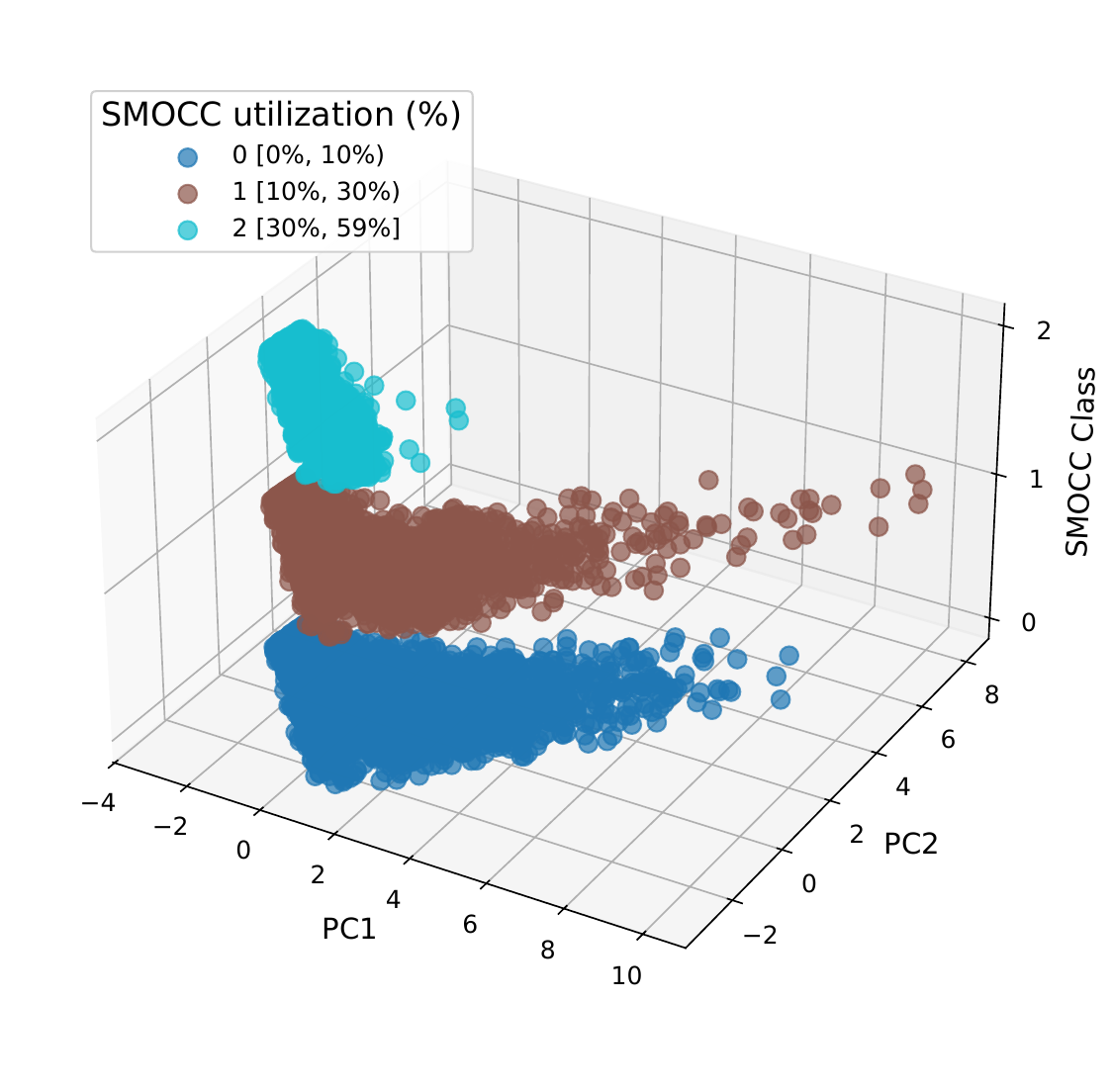}
        \caption{CNN SMOCC}
        \label{fig:CNN_SMOCC_pca3d}
    \end{subfigure}
    \begin{subfigure}[b]{0.32\textwidth}
        \centering
        \includegraphics[width=\linewidth, trim={0 0 0 0}, clip]{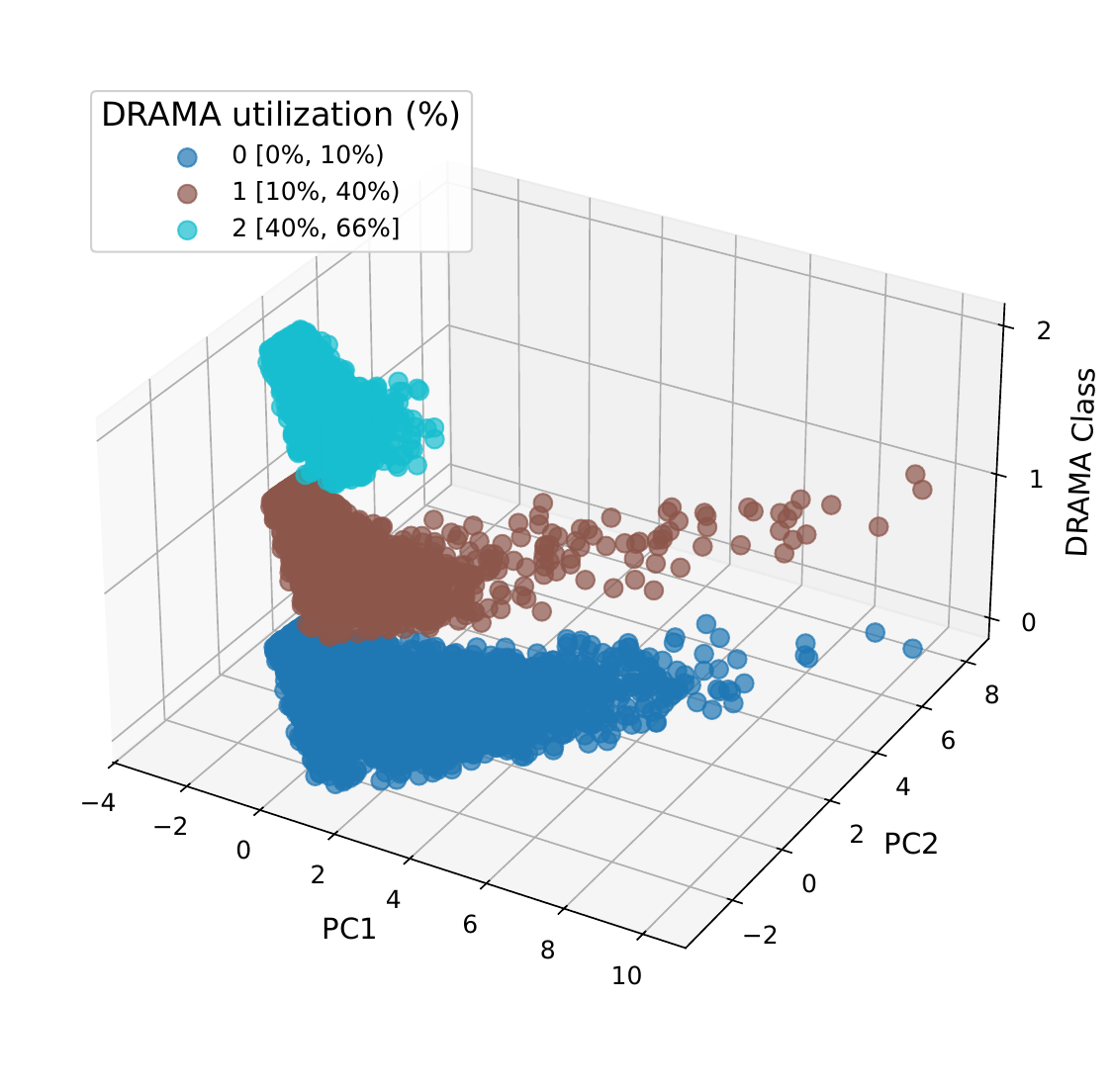}
        \caption{CNN DRAMA}
        \label{fig:CNN_DRAMA_pca3d}
    \end{subfigure}
    \begin{subfigure}[b]{0.32\textwidth}
        \centering
        \includegraphics[width=\linewidth, trim={0 0 0 0}, clip]{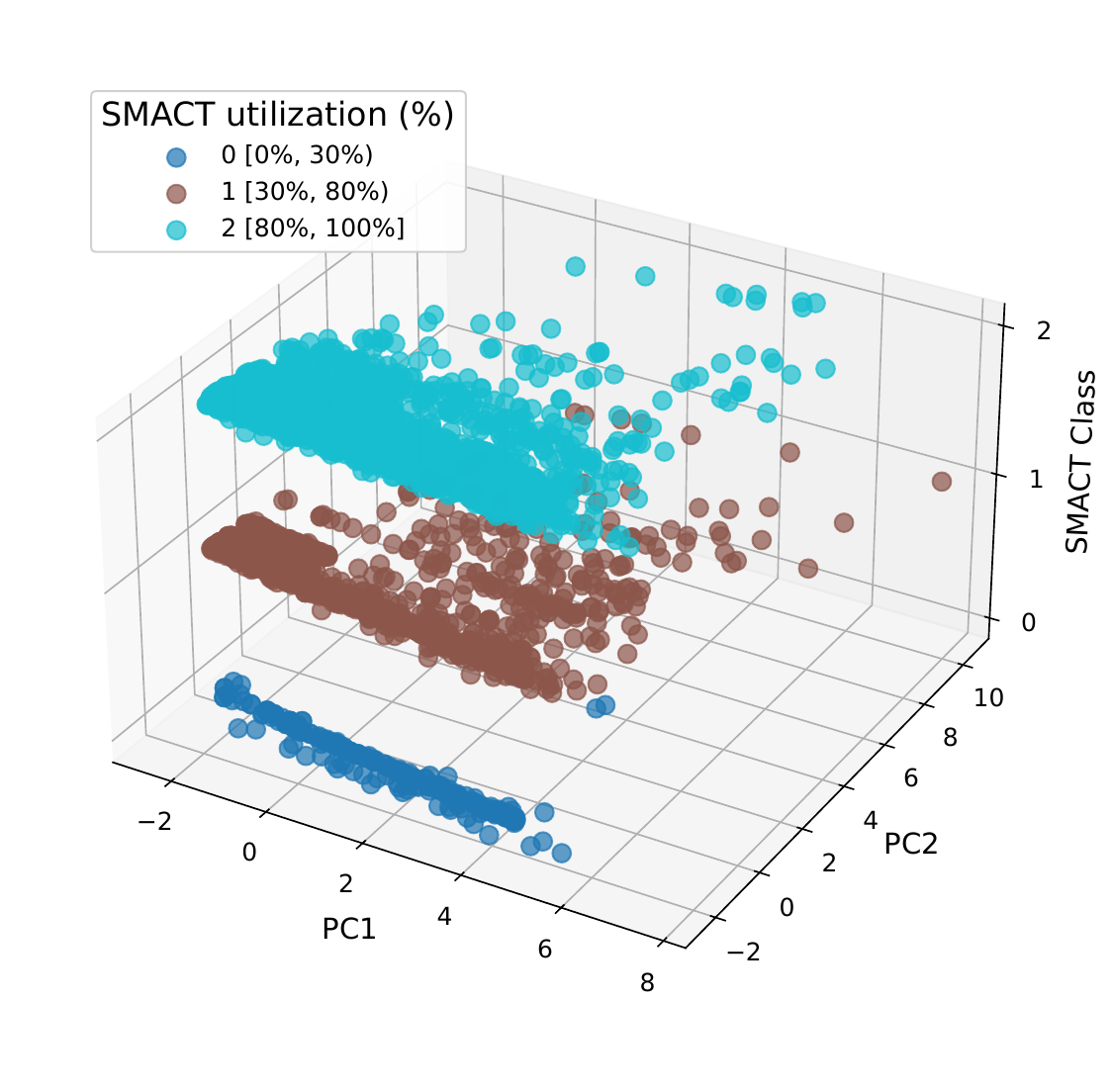}
        \caption{Transformer SMACT}
        \label{fig:Trans_SMACT_pca3d}
    \end{subfigure}
    \begin{subfigure}[b]{0.32\textwidth}
        \centering
        \includegraphics[width=\linewidth, trim={0 0 0 0}, clip]{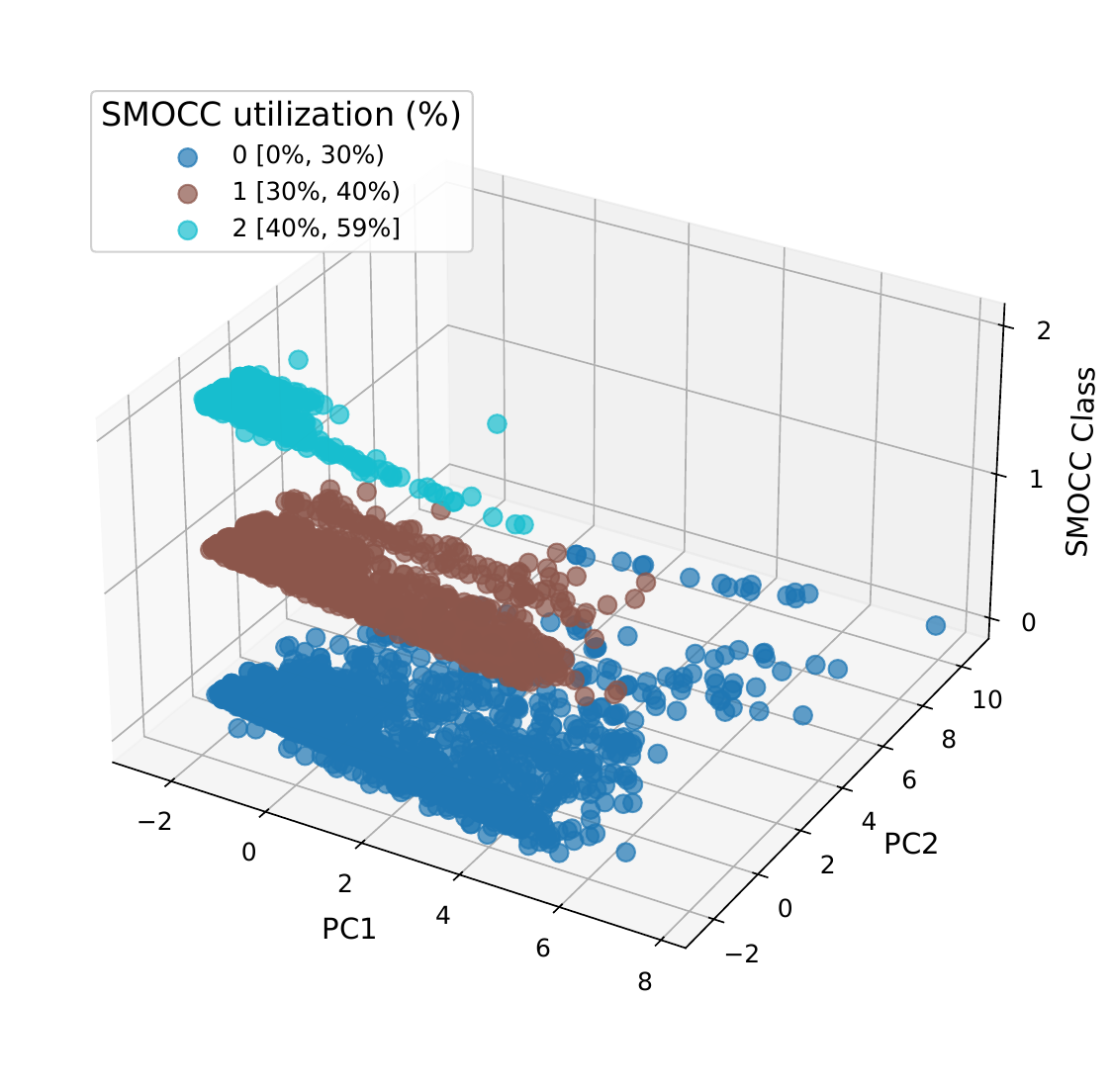}
        \caption{Transformer SMOCC}
        \label{fig:Trans_SMOCC_pca3d}
    \end{subfigure}
    \begin{subfigure}[b]{0.32\textwidth}
        \centering
        \includegraphics[width=\linewidth, trim={0 0 0 0}, clip]{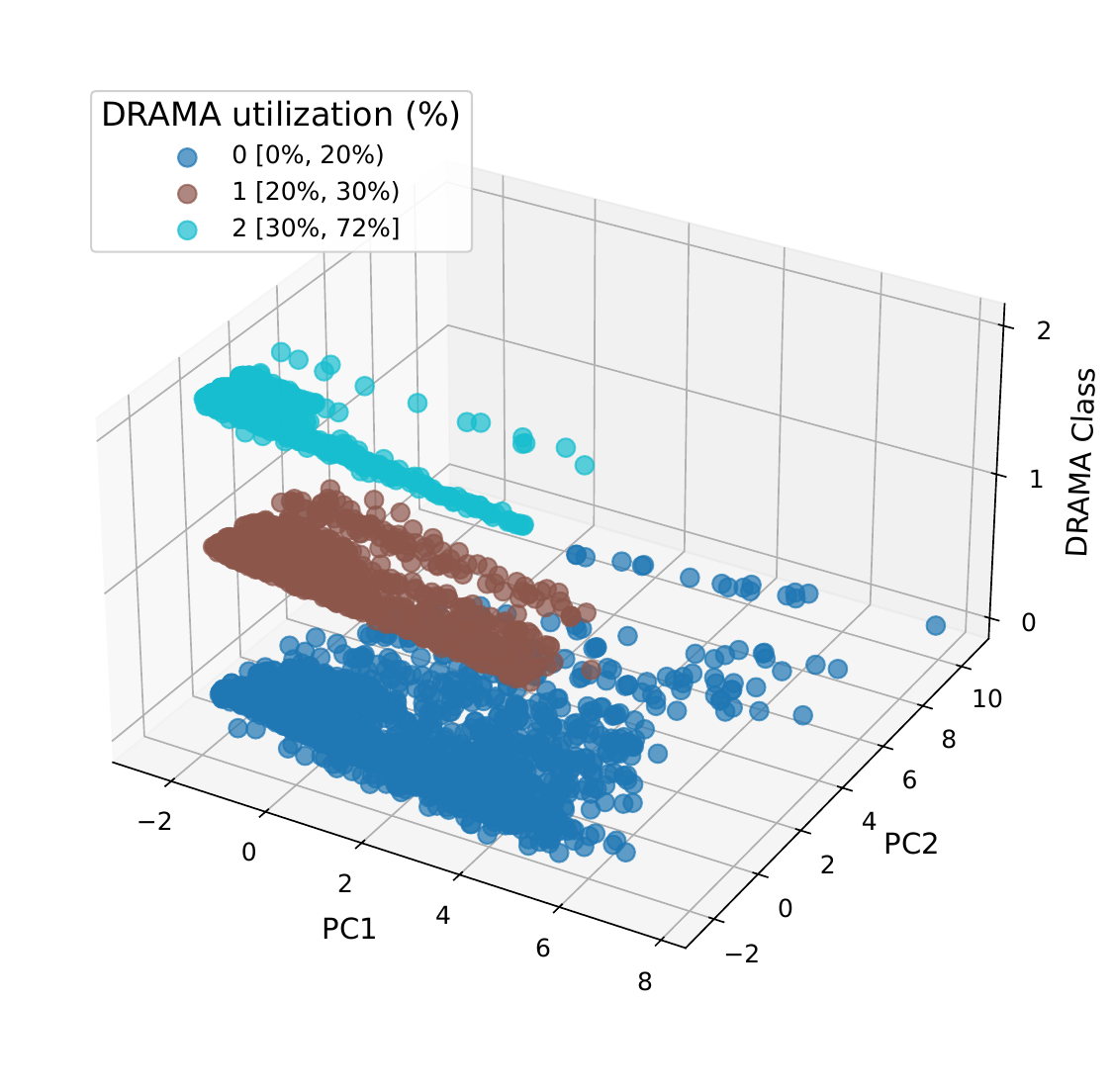}
        \caption{Transformer DRAMA}
        \label{fig:Trans_DRAMA_pca3d}
    \end{subfigure}
    \caption{Principal Component Analysis (PCA) of the datasets across different neural network architectures for utilization metrics.}
    \label{fig:dataset_PCA_util}
\end{figure}

\Cref{tbl:utilization_results} in Appendix reveals clear performance disparities across architectures, with accuracy degrading substantially for non-MLP workloads — DRAMA scores as low as 62\% on Transformers. Recall, critical for correctly identifying high-utilization workloads, remains poor across most models. As shown in \Cref{fig:dataset_PCA_util} (Appendix) and discussed in \Cref{subsec:classification_vs_regression}, utilization classes lack clear separability, with CNN and Transformer data points overlapping across all classes, preventing models from learning meaningful decision boundaries. A further limitation is that using single-workload predictions for collocation can implicitly assume additive utilization across workloads. In practice, utilization is not additive because co-located workloads exhibit non-linear interactions arising from shared GPU resource contention and differing execution phases.
\section{Discussion and Limitations}
\label{sec:discussion_limitations}
\textbf{Overcoming Limitations of Existing Estimators.} Current estimation tools often suffer from intrusive designs that require deep integration into upper software layers, making them impractical for general use. Furthermore, many existing solutions struggle with high latency overhead on the scheduler’s critical path, frequent misestimations, and inability to generalize across different models and GPU generations.

\textbf{Lessons from Initial Experimental Failures.} Early attempts using simple MLP datasets revealed that basic models are insufficient for robust prediction. A discrete classifier achieved only $\sim$69\% accuracy due to class imbalances, while an \textit{ExtraTreeRegressor} showed poor generalization with a $\pm$1.2 GB error margin, primarily because it overfitted to a narrow feature set that did not adhere to the principles in \Cref{subsubsec:dataset_consideration}.

\textbf{Current Constraints of the GPUMemNet Framework.} While \textbf{GPUMemNet} is effective within its scope, it currently relies on \texttt{torchsummary}, which may struggle with custom or non-standard layers. Its performance is tied to training data diversity; without representative data for components like \texttt{Conv1D} or residual connections, or support for model/pipeline parallelism, its applicability remains limited to data-parallel setups.

\textbf{The Challenges of Estimating GPU Utilization.} GPU utilization estimation remains an understudied and challenging area where our initial attempts were not successful. Using utilization as a proxy for resource interference is not straightforward because these metrics are not additive and vary across GPU architectures, suggesting that future work must address these non-linear hardware utilization characteristics.

\textbf{Defining the Path Toward Non-Intrusive Integration.} A practical estimator should remain decoupled from application-level code, ML-framework logic, and the resource manager’s scheduling or mapping path, so that online placement decisions are not delayed by estimator-specific integration. By non-intrusive, we mean that developers should not need to modify training code, manually extract model summaries, or place such information in predefined locations, and resource managers should not need to invoke estimator-specific logic at runtime. Instead, resource metadata should be generated earlier in the workflow—for example during model development, testing, or packaging—and made available before the workload reaches the resource manager.

\section{Conclusion}
\label{sec:conclusion}

In this paper, we investigated GPU memory and utilization estimation, providing aqualitative and quantitative analysis of their respective trade-offs. We presented a data-driven, extensible framework designed to estimate resource requirements for deep learning training tasks. By addressing the practical limitations of current estimation methods, more informed and efficient resource management within GPU clusters can be achieved.

\bibliographystyle{ACM-Reference-Format}
\bibliography{references}

\clearpage
\onecolumn 

\FloatBarrier
\clearpage
\section{Appendix}
\label{appendix}

\begin{table}[t]
\centering
\caption{Parameter ranges for the MLP dataset.}
\begin{tabularx}{0.6\columnwidth}{|l|X|}
\hline
\textbf{Parameter} & \textbf{Range} \\ \hline
Input Size & 4 to 4,096 \\ \hline
Output Size & 1 to 1,024 \\ \hline
Linear Layers &  2 to 12 \\ \hline
Batch Normalization Layers & 0 to 11\\ \hline
Dropout Layers & 0 to 11 \\ \hline
Batch Size & 4 to 1,024 \\ \hline
Activation Function & ReLU, ELU, Tanh, etc. \\ \hline
Architecture & pyramid, uniform, bottleneck, gradual \\ \hline
\end{tabularx}
\label{tab:mlp_parameters}
\end{table}

\begin{table}[t]
\centering
\caption{Parameter ranges for the CNN dataset.}
\begin{tabularx}{0.6\columnwidth}{|l|X|}
\hline
\textbf{Parameter} & \textbf{Range} \\ \hline
Input Size & 32x32 to 224x224 \\ \hline
Input Channels & 1 to 3 \\ \hline
Output Size & 2 to 22,000 \\ \hline
Filters & 16 to 512 with sizes of 3×3, 2x2, 1x1 \\ \hline
Conv2d Layers & 1 to 29 \\ \hline
BatchNorm2d Layers & None to 29 \\ \hline
Dropout Layers & None to 29 \\ \hline
Linear Layers & 1 to 1 \\ \hline
Batch Size & 2 to 62 \\ \hline
Activation Function & ReLU, ELU, Tanh, etc. \\ \hline
Architecture & pyramid, uniform, bottleneck, gradual, etc. \\ \hline
\end{tabularx}
\label{tab:cnn_subdataset}
\end{table}

\begin{table}[t]
\centering

\caption{Parameter ranges for the collected Transformers dataset.}
\begin{tabularx}{0.6\columnwidth}{|l|X|}
\hline
\textbf{Parameter} & \textbf{Range} \\ \hline

Input Size & 50,000, 1,000,000 \\ \hline
Sequence Length & 128 to 1024 \\ \hline
Embedding Size & 128 to 2048 \\ \hline
Number of classes & 400 to 1000 \\ \hline
Linear Layers & 9 to 157 \\ \hline
LayerNorm Count & 4 to 78 \\ \hline
Dropout Layers & 6 to 117 \\ \hline
Batch Size & 1 to 128 \\ \hline
Architecture & standard, decoder only, hybrid \\ \hline
\end{tabularx}
\label{tab:transformer_parameters}

\end{table}

\begin{table}[t]
\centering
\caption{GPU Memory and extracted total parameters and total activations of datasets}
\begin{tabularx}{0.6\columnwidth}{|l|X|}
\hline
\textbf{Parameter} & \textbf{Range} \\ \hline

\multicolumn{2}{|c|}{\textbf{MLP dataset}} \\ \hline \hline
Total Parameters & 27 to 159,856,482 \\ \hline
Total Activations & 10 to 148,066 \\ \hline
GPU Memory Need & 1,443 to 4,925 MB \\ \hline

\multicolumn{2}{|c|}{\textbf{CNN dataset}} \\ \hline \hline

Total Parameters & 704 to 329,307,377 \\ \hline
Total Activations & 24,514 to 5,317,481,490 \\ \hline
GPU Memory Need & 1,703 to 40,369 MB \\ \hline

\multicolumn{2}{|c|}{\textbf{Transformer dataset}} \\ \hline \hline
Total Parameters & 7,292,531 to 3,105,662,679 \\ \hline
Total Activations & 19,184 to 3,194,470 \\ \hline
GPU Memory Need & 1,683 to 40,369 MB \\ \hline
\end{tabularx}
\label{tab:target_extracted_parameters}
\end{table}

\begin{table}[t]
\centering
\caption{Accuracy results and recall on high utilization classes for the GPU hardware metric utilization estimations with MLP-based models.}
\begin{tabularx}{0.6\columnwidth}{|l|c|X|c|c|}
\hline
\textbf{Dataset} & \textbf{Metric} & \textbf{Bins (low, medium, high)} &
\textbf{Acc.} & \textbf{Recall}\\
\hline \hline
MLP 
    & \makecell[l]{SMACT} 
    & [0.0, 0.2), [0.2, 0.7), [0.7, 1.0] & 0.92 & 0.79\\
\cline{2-5}
    & \makecell[l]{SMOCC} 
    & [0.0, 0.1), [0.1, 0.3), [0.3, 1.0] & 0.94 & 0 \\
\cline{2-5}
    & \makecell[l]{DRAMA} 
    & [0.0, 0.2), [0.2, 0.7), [0.7, 1.0] & 0.88 & 0.27 \\
\hline

\makecell[l]{CNN}
    & \makecell[l]{SMACT}
    & [0.0, 0.3), [0.3, 0.9), [0.9, 1.0] & 0.77 & 0.87 \\
\cline{2-5}
    & \makecell[l]{SMOCC} 
    & [0.0, 0.1), [0.1, 0.3), [0.3, 1.0] & 0.71 & 0.88 \\
\cline{2-5}
    & \makecell[l]{DRAMA} 
    & [0.0, 0.3), [0.3, 0.9), [0.9, 1.0] & 0.73 & 0.88 \\
\hline

\makecell[l]{Transformer}
    & \makecell[l]{SMACT} 
    & [0.0, 0.3), [0.3, 0.8), [0.8, 1.0] & 0.82 & 0.89 \\
\cline{2-5}
    & \makecell[l]{SMOCC} 
    & [0.0, 0.3), [0.3, 0.4), [0.4, 1.0] & 0.68 & 0.36 \\
\cline{2-5}
    & \makecell[l]{DRAMA} 
    & [0.0, 0.3), [0.3, 0.8), [0.8, 1.0] & 0.62 & 0.74 \\ 
\hline

\end{tabularx}

\label{tbl:utilization_results}
\end{table}


\end{document}